\documentclass[aps,prb,reprint,superscriptaddress,citeautoscript]{revtex4-1}

\usepackage{amsmath}
\usepackage{amsfonts}
\usepackage{amssymb}
\usepackage{graphicx}
\usepackage{times}

\graphicspath{{.}{./EPS/}}

\usepackage{bm}
\newcommand* {\vek}[1]{{\ensuremath{\bm{\mathrm{#1}}}}}
\newcommand* {\ee}{\ensuremath{\mathrm{e}}}

\begin{document}

\title{Refraction in spacetime}

\author{Markku J\"{a}\"{a}skel\"{a}inen}
\affiliation{Institute of Fundamental Sciences and MacDarmid Institute for Advanced
Materials and Nanotechnology, Massey University, Manawatu Campus, Private Bag
11~222, Palmerston North 4442, New Zealand}

\author{Marijke Lombard}
\affiliation{Institute of Fundamental Sciences and MacDarmid Institute for Advanced
Materials and Nanotechnology, Massey University, Manawatu Campus, Private Bag
11~222, Palmerston North 4442, New Zealand}

\author{Ulrich Z\"{u}licke}
\affiliation{Institute of Fundamental Sciences and MacDarmid Institute for Advanced
Materials and Nanotechnology, Massey University, Manawatu Campus, Private Bag
11~222, Palmerston North 4442, New Zealand}
\affiliation{Centre for Theoretical Chemistry and Physics, Massey University,
Albany Campus, Private Bag 102~904, North Shore MSC, Auckland 0745, New
Zealand}

\date{\today}

\begin{abstract}
Refraction, interference, and diffraction serve as distinguishing features for
wave-like phenomena. While they are normally associated only with a purely
spatial wave-propagation pattern, analogs to interference and diffraction involving
the spatio-{\em temporal\/} dynamics of waves in one dimension (1D) have been
pointed out. Here we complete the triplet of analogies by discussing how
spatio-temporal analogs to \emph{refraction\/} are exhibited by a quantum
particle in 1D that is scattering off a step barrier. Similarly, birefringence in
spacetime occurs for a spin-1/2 particle in a magnetic field. These examples
serve to illustrate basics of quantum time evolution from a new perspective.
\end{abstract}

\maketitle

\section{Introduction}

Quantum mechanics is essentially a wave theory, and analogs with classical optics
can be fruitful for providing a deeper understanding of quantum phenomena, both
for pedagogical purposes and for more research-oriented explorations. The direct
analogy between the Helmholtz equation and the stationary Schr\"{o}dinger
equation leads to the existence of identical phenomena in classical optics and
quantum mechanics for purely static situations~\cite{Brukner1997}. The three
archetypical wave-like phenomena encountered in undergraduate textbooks are
diffraction around sharp corners, interference between waves propagating along
different paths, and refraction at boundaries between regions with different wave
speeds.

As recent years have seen a growing interest in quantum dynamics,
time-dependent treatments of quantum mechanics have found their way into
textbooks~\cite{Robinett2006,Tannor2007,Thaller2000}. In this context, it is
worthwhile to note that computerised studies of wave-packet scattering were
pioneered in this journal already in 1967\cite{Goldberg1967} and have been the
topic for research ever since, revealing  and visualising nonintuitive features of
simple systems~\cite{Bramhall1970,Kiriuscheva1998}.

For quantum dynamics in one spatial dimension, an exact analogy with stationary
diffraction in optics was discovered by Moshinsky in 1952 in his work on diffraction
in time~\cite{Moshinsky1952}; see also Ref.~\onlinecite{Brukner1997}. Since then,
the field has expanded considerably~\cite{delCampo2009}. Experiments with
ultracold atoms have observed interference in time~\cite{Szriftgiser1996} and, more
recently, interference in time without spatial dynamics has been shown to give rise
to interference effects in the time-energy domain~\cite{Lindner2005}.

In this paper, we complete the set of spatio-temporal analogues of wave
phenomena by pointing out the occurrence of refraction in spacetime for wave-packet
propagation through a one-dimensional (1D) inhomogeneous medium. As a further
application, the case of birefringence~\cite{birefringNote} is considered whose
spatio-temporal analog appears in the propagation of a wave packet that is an
equal superposition of  spin-up and spin-down states through a region with spatially
inhomogeneous magnetic field.

The remainder of this article is organized as follows. In the following
Sec.~\ref{sec:ordRef}, we provide background information and establish notational
details by discussing ordinary (stationary) refraction in space for the solutions of
electromagnetic and Schr\"odinger wave equations. The emergence and properties
of refraction in spacetime are explained based on a ray picture in Sec.~\ref{sec:refInT}.
Quantum simulations of wave-packet dynamics are presented in Sec.~\ref{sec:quantsim}
to show the validity of the basic ray model for refraction in spacetime, as well as to
illustrate regimes where it needs to be modified/breaks down. Section~\ref{sec:biRef}
is devoted to the phenomenon of birefringence exhibited by spin-1/2 particles entering
a region of finite and spatially uniform magnetic field from a field-free region. Further
discussion and our conclusions are presented in Sec.~\ref{sec:discConc}.

\section{Ordinary refraction of classical and quantum-probability waves}
\label{sec:ordRef}

Each Cartesian component of the electromagnetic field in a dielectric medium
satisfies a wave equation whose solutions can be most generally expressed as
a superposition of single-frequency wave amplitudes $\Psi_{\text{el}} (\vek{x},t)
=\psi_{\text{el}}(\vek{x}) \, \ee^{-i\omega t}$. The stationary wave amplitude
is found from the Helmholtz equation
\begin{equation}\label{newcwe}
\vek{\nabla}^2 \psi_{\text{el}} (\vek{x}) + k^2 \psi_{\text{el}} (\vek{x})=0 \quad ,
\end{equation}
where $k$ denotes the wave number. In vacuum, $k=k_0 \equiv \omega/c_0$,
where $c_0 = 1/\sqrt{\epsilon_0\mu_0}$ is the speed of light in vacuum. The effect
of a medium is described by the refractive index $n_{\text{el}} = k/k_0$ which, in
general, is a frequency-dependent materials parameter. 

For stationary situations, the optical case is analogous to quantum mechanics.
Consider the time-dependent Schr\"{o}dinger equation for a massive point-particle
subject to an external potential $V(\vek{x})$, which is given by
\begin{equation}\label{tdse}
i\hbar\frac{\partial}{\partial t}\Psi_{\text{qu}} (\vek{x},t) = -\frac{\hbar^2}{2m}
\vek{\nabla}^2 \Psi_{\text{qu}} (\vek{x},t) + V(\vek{x}) \, \Psi_{\text{qu}} (\vek{x},t)
\quad .
\end{equation}
Similar to the electromagnetic case, we can make a separation \textit{ansatz\/}
$\Psi_{\text{qu}}(\vek{x},t)=\psi_{\text{qu}}(\vek{x}) \, \ee^{-iE t/\hbar}$ and find the
time-independent Schr\"{o}dinger equation
\begin{equation}\label{dltise}
-\frac{\hbar^2}{2m} \vek{\nabla}^2\psi_{\text{qu}} (\vek{x})+V(\vek{x})\, \psi_{\text{qu}}
(\vek{x})=E \, \psi_{\text{qu}} (\vek{x}) \quad .
\end{equation}
Rearranging yields
\begin{equation}
\vek{\nabla}^2 \psi_{\text{qu}} (\vek{x}) + \frac{2m}{\hbar^2}\left[ E-V(\vek{x})\right]
\psi_{\text{qu}}(\vek{x})=0 \quad ,
\end{equation}
which has the same general form as the classical Helmholtz equation \eqref{newcwe}.
The formal analogy\cite{Bunge1967,BailerJones2009} with the optical case is complete
for a constant potential $V(\vek{x})\equiv V_0\le E$. This can be made explicit by 
defining $k=\sqrt{2m(E-V_0)/\hbar^2}$ and $\omega = E/\hbar$. (We consider here only
propagating solutions corresponding to real $k$.) Note that the potential $V_0$ determines
the wave number, and the quantity
\begin{equation}\label{eq:quantN}
n_{\text{qu}} = \sqrt{1 - \frac{V_0}{\hbar\omega}}
\end{equation}
suggests itself as the quantum-mechanical equivalent of the refractive index in
optics~\cite{refIndexNote}. With this definition of $n_{\text{qu}}$, the solutions for
classical optics will be identical to the ones for quantum optics whenever
$n_{\text{el}}$ is identical to $n_{\text{qu}}$.

\begin{figure}[t]
\begin{center}
\includegraphics[width=8cm]{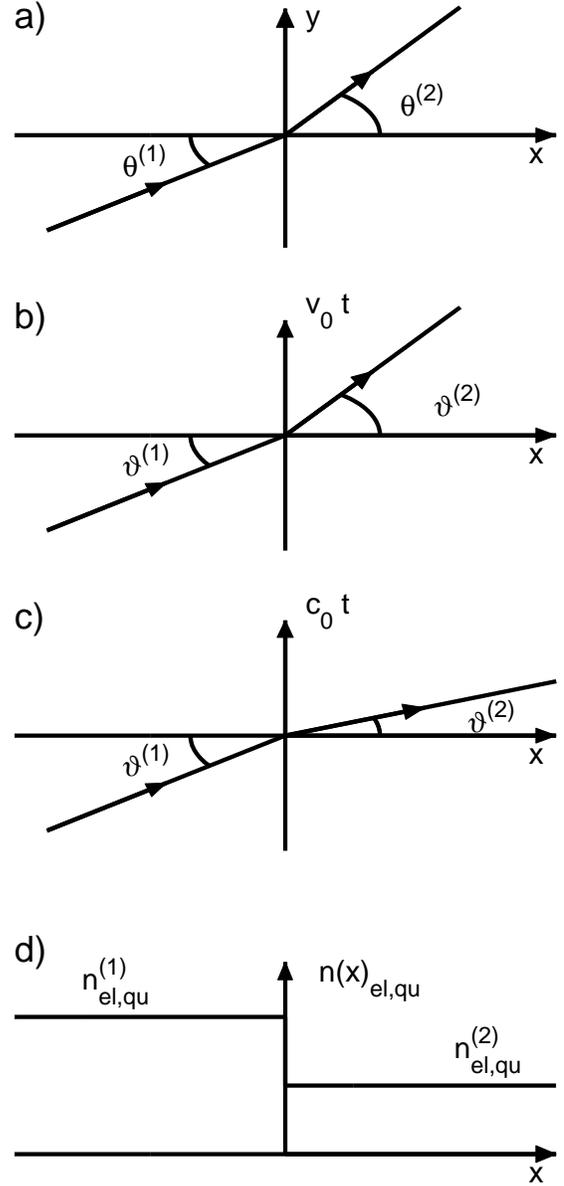}
\caption{\label{fig:sine}
a)~Path taken by an optical or quantum wave through two adjoining media with
different refractive indices $n^{(1)}$ and $n^{(2)}$. b)~Trajectory of a quantum
wave packet in $1+1$-dimensional spacetime, with time measured in terms of
propagation distance. The wave packet exhibits a refraction-like phenomenon
with the refraction angles satisfying a modified Snell's law relation [given by
Eq.~(\ref{eq:snellTqu})]. c)~Trajectory of an optical wave in $1+1$-dimensional
spacetime when crossing between two nondispersive media. The Snell's-law
equivalent describing its refraction in spacetime is different
[see Eq.~(\ref{eq:snellTel})]. d)~ Refractive index as a function of spatial coordinate.}
\end{center}
\end{figure}

The Helmholtz equation has general plane-wave solutions of the form
$\psi_{\text{el,qu}} = \psi_{\text{el,qu}}^{(0)} \, \ee^{i \vek{k}\cdot\vek{x}}$. Here
$\vek{k}$ denotes the wave vector with magnitude $|\vek{k}| = k$ and direction
$\hat{\vek{k}}$ coinciding with the propagation direction of the wave, and
$\psi_{\text{el,qu}}^{(0)}$ is a constant. When such a plane wave is incident at a
nonzero angle on the interface between two media with different refractive indices,
a change of direction is observed for the part of the wave passing from one
medium into the other. This phenomenon of \emph{refraction\/} is illustrated in
Fig.~\ref{fig:sine}~a). As both the wave frequency and wave-vector component
parallel to the interface are conserved in this process, the relation
\begin{equation}\label{eq:SnellK}
k^{(1)} \sin{\theta^{(1)}} = k^{(2)} \sin{\theta^{(2)}}
\end{equation}
holds. In terms of the refractive indices in the two adjoining spatial regions,
Eq.~(\ref{eq:SnellK}) can be written as
\begin{equation}\label{snelloptics}
n_{\text{el,qu}}^{(1)}\,\sin{\theta^{(1)}}=n_{\text{el,qu}}^{(2)}\,\sin{\theta^{(2)}} \quad .
\end{equation}
As we consider electromagnetic waves as the example for the classical case, the
refractive index is a property of the dielectric media occupying spatial regions
$x>0$ and $x<0$, respectively. For the quantum case, refractive indices are
defined in accordance with Eq.~(\ref{eq:quantN}) for the situation of piecewise
constant external potential
\begin{equation}\label{PotentialStep}
V(x) =
\begin{cases}
V_0^{(1)} & \text{$x < 0$} \\
V_{0}^{(2)} & \text{$x > 0$}
\end{cases} \quad .
\end{equation}

Equation~(\ref{snelloptics}) is the familiar Snell's law for refraction, which also
follows from the principle of least action giving the shortest time between two
spatial points. The situation with $\theta^{(1)} < \theta^{(2)}$ shown in
Fig.~\ref{fig:sine}~a) corresponds to $n^{(1)} > n^{(2)}$ as illustrated in
Fig.~\ref{fig:sine}~d). In quantum mechanics, this scenario would be realized by
a potential step having $V_0^{(1)} < V_0^{(2)}$. In a general scattering
situation, only part of the wave will be transmitted through the interface and
undergo refraction, while the other part will be reflected back into the medium
from where the wave is incident. To properly describe both reflection and refraction
at an interface, the Fresnel equations have to be used\cite{BornWolf1999}.

\section{Refraction in 1+1 dimensions}
\label{sec:refInT}

We now consider the space-time description of rays in an inhomogeneous medium
in one spatial dimension and elucidate the phenomenon of refraction \emph{in
spacetime\/}. In order to define angles and trigonometric functions in the plane
spanned by time and the spatial coordinate, these two quantities need to have
identical dimensions. The simplest way to ensure this is to multiply time by a constant
reference speed, turning it into an equivalent distance, $t\rightarrow v_0 t$. For the
case of optical waves, a natural choice is $v_{0,\text{el}} = c_0$, the speed of light
in vacuum, thus giving the time axis the meaning of propagated distance in empty
space. In the same spirit, we use $v_{0,\text{qu}} = \sqrt{2 E/m}$ for quantum-probability
waves.

In a homogeneous 1D medium, the speed of a wave packet is its local group
velocity $v_{\text{g}}=\partial\omega/\partial k$, and the slope of its worldline in the
$v_0 t$--$x$ diagram is given by $\Delta (v_0 t)/\Delta x\equiv \tan\vartheta$. 
Here we use $\vartheta$ to denote the angle in the spacetime diagram, as opposed
to $\theta$ used for spatial refraction. The universal flow of time in two adjoining regions
1 and 2 where the group velocity is different implies occurrence of a refraction
phenomenon in the $v_0 t$--$x$ space described by the relation between incident
and transmitted angles [see Fig.~\ref{fig:sine}(b)]
\begin{equation}\label{eq:SnellV}
v_{\text{g}}^{(1)} \tan\vartheta^{(1)} = v_{\text{g}}^{(2)} \tan\vartheta^{(2)} \quad .
\end{equation}
Inserting the definition of the refractive index into the formula for the group velocity
yields the general relation
\begin{equation}\label{eq:genGroupV}
v_{\text{g}} = \frac{v_0}{n} \left( 1 + \alpha_v \, \frac{\omega}{n}\frac{\partial n}
{\partial \omega} \right)^{-1} \quad ,
\end{equation}
where $\alpha_v = v_0 k_0/\omega$ is the ratio of group and phase velocities in
the medium with $n=1$ (i.e., vacuum). Combining Eqs.~(\ref{eq:SnellV}) and
(\ref{eq:genGroupV}), we find
\begin{eqnarray}\label{eq:genTimeSnell}
&& n^{(2)} \left( 1 + \alpha_v \, \frac{\omega}{n^{(2)}}\frac{\partial n^{(2)}}{\partial
\omega}\right) \tan\vartheta^{(1)} \nonumber \\ && \hspace{2cm} = n^{(1)} \left(
1 + \alpha_v \, \frac{\omega}{n^{(1)}} \frac{\partial n^{(1)}}{\partial \omega} \right)
\tan\vartheta^{(2)}
\end{eqnarray}
as the Snell's-law-type relation describing refraction in spacetime. For the simple
case of an electromagnetic wave crossing between nondispersive media (i.e., for
constant $n_{\text{el}}^{(1,2)}$),
Eq.~(\ref{eq:genTimeSnell}) specializes to
\begin{subequations}
\begin{equation}\label{eq:snellTel}
n_{\text{el}}^{(2)} \tan{\vartheta_1} = n_{\text{el}}^{(1)} \tan{\vartheta_2} \quad .
\end{equation}
This situation is depicted in Fig.~1~c). A different result is obtained in the quantum
case where the refractive index is frequency-dependent [cf.\ Eq.~(\ref{eq:quantN})],
yielding
\begin{equation}\label{eq:snellTqu}
n_{\text{qu}}^{(1)} \tan{\vartheta_1} = n_{\text{qu}}^{(2)} \tan{\vartheta_2} \quad .
\end{equation}
\end{subequations}
The ray picture for this case is illustrated in Fig.~1~b).

We have found a refraction-like phenomenon in the spatio-temporal evolution of a
wave propagating in one spatial dimension. The kinematic differences associated
with refraction in spacetime compared to ordinary refraction (in space) are captured in
Eqs.~(\ref{eq:SnellV}) and (\ref{eq:SnellK}). Snell's law (\ref{snelloptics}) universally
holds for refraction in space for any type of waves because refractive indices are
sensibly defined in terms of the wave number (or phase velocity) and not group
velocity. It is, however, the latter that determines refraction in spacetime, hence the
$1+1$-dimensional equivalent of Snell's law [Eq.~(\ref{eq:genTimeSnell})] is not simply
determined by the values of refractive indices but also influenced by their frequency
dependence. As a result, spacetime-refracted beams can be bent in opposite directions
for waves that are subject to the same spatial profile of the refractive index, as is
illustrated in Figs.~\ref{fig:sine}~b) and \ref{fig:sine}~c).

So far, we have stressed the wave properties and refraction angles in spacetime,
without referring to the associated particle properties exhibited in scattering events.
Within the particle picture, the slope of the worldline in a spacetime diagram corresponds
to the particle's velocity. Thus in the situation depicted in Fig.~\ref{fig:sine}~b), which
applies to refraction with $\vartheta_2 > \vartheta_1$, the particle speeds up after the
discontinuity~\cite{note:phaseVSgroup}.

\section{Refraction in spacetime exhibited in simulated quantum wave-packet dynamics}
\label{sec:quantsim}

In the previous section, we considered plane-wave solutions to the stationary (i.e.,
time-independent) wave equations, both Helmholtz's and Schr{\"o}dinger's. To make
our results more visualizable, we now turn to the dynamics of quantum-mechanical
wave packets in a single spatial dimension. For a dynamical situation, we can
superimpose a continuum of scattering solutions to construct the wave function
corresponding to any given sufficiently smooth initial state. As a model system, we
choose the textbook example of scattering off a finite potential step, given by
Eq.~(\ref{PotentialStep}) with $V_0^{(1)}=0$ and $V_0^{(2)}=V_0$. Assuming the
wave to be incident from region 1, the stationary wave functions for the situation at
hand are
\begin{equation}
\psi_k (x) =
\begin{cases}
\ee^{i k x} + r(k) \, \ee^{-i k x} & \text{$x < 0$} \\
 t(k) \, \ee^{i k^\prime x} & \text{$x > 0$}
\end{cases} \quad .
\end{equation}
Here $k=\sqrt{2 m E/\hbar^2}$, $k^\prime=\sqrt{2 m (E - V_0)/\hbar^2}$, and the
reflection and transmission amplitudes are (see, e.g.,
Ref.~\onlinecite{Robinett2006})
\begin{equation}
r(k)=  \Theta(k)\, \frac{k-k^\prime}{k+k^\prime} \,\, , \quad
t(k)=\Theta(k)\, \frac{2\sqrt{k k^\prime}}{k+k^\prime} \,\, ,
\end{equation}
with $\Theta(k)$ denoting Heaviside's step function. As a free Gaussian wave
packet centered at $x=x_0$ and evolving in time can be expressed as a linear
combination of plane waves,
\begin{equation}
\psi_{\text{G}}(x,t)=\int_{-\infty }^{ \infty} dk \,\,\,\, \ee^{-ik (x - x_0)} \,
\ee^{ -\frac{(k-k_{0})^2}{2 \Delta k^2}} \, \ee^{-i\frac{\hbar k^2}{2m}t} \, ,
\end{equation}
its dynamics when scattering off a potential step is given by the analogous
superposition of solutions $\psi_k$:
\begin{equation}
\psi(x) = \int_{-\infty }^{ \infty} dk \,\,\,\, \psi_k (x)\, \ee^{ i k x_0} \,
\ee^{ -\frac{(k-k_{0})^2}{2 \Delta k^2}} \, \ee^{-i\frac{\hbar k^2}{2m}t} \, .
\end{equation}

\begin{figure}[t]
\begin{center}
\includegraphics[width=8cm]{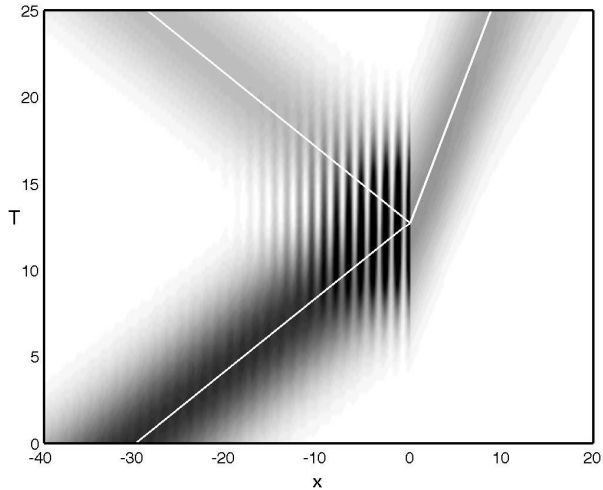}
\caption{\label{fig:Refraction}
Refraction in spacetime for a Gaussian wave packet. The coordinate $x$ is measured
in terms of an arbitrary unit of length $\ell$, and $T=\hbar t/(m\ell^2)$. Parameters used
in the simulation: $\Delta k =0.1/\ell $, $x_{0}=-30\,\ell$, $k_{0}=2.35/\ell$, and $V_{0}=
2.5\, \hbar^2/(m\ell^2)$. Note that the wave packet is refracted away from the normal as
a result of the positive potential step. The white lines are plots of the trajectories for the
incident, reflected and the refracted rays obeying the equivalent of Snell's law for
refraction in spacetime [Eq.~(\ref{eq:snellTqu})].}
\end{center}
\end{figure}
In Fig.~\ref{fig:Refraction}, we show the simulated dynamics of a Gaussian wave
packet scattering off a potential step, with worldlines for the incoming, reflected, and
refracted trajectories superimposed. The parameters used in this calculation were
chosen such that $|r(k)|^2\approx 0.5$, which ensures sufficient visibility of the 
fringes arising from interference between incoming and reflected parts of the wave
packet. The transmitted part is seen to slow down as the scattering occurs against
a positive potential step. In terms of our discussion of refraction in spacetime,
Fig.~\ref{fig:Refraction} corresponds to the scenario illustrated in
Fig.~\ref{fig:sine}~b), with the worldline of the transmitted beam being refracted away
from the normal as scattering for the quantum wave occurs against a medium with
lower index of refraction.

\begin{figure}[t]
\begin{center}
\includegraphics[width=8cm]{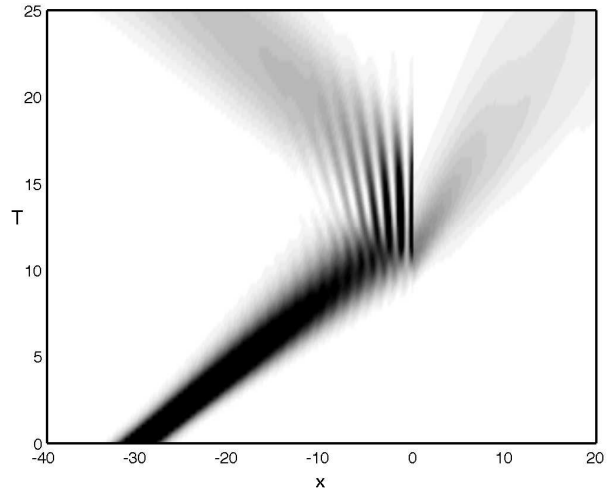}
\caption{\label{fig:ComboPlot}
Refraction in spacetime for a Gaussian wave packet that broadens significantly
during the scattering event. Parameters are the same as in Fig.~\ref{fig:Refraction},
except that $\Delta k =0.5/\ell $. The interference fringes between the incident and
reflected components are seen to be curved here, in contrast to the phase fronts
in Fig.~\ref{fig:Refraction} which are straight.}
\end{center}
\end{figure}

For the ray picture to properly describe the propagation of a wave packet, the latter
should not spread appreciably during the scattering process to avoid effects due to
broadening. Free wave packets broaden due to dispersive effects from the initial
distribution of momenta. This will become noticeable over a time scale
\begin{equation}\label{BroadeningTime}
t_{\text{broad}} \approx \frac{m\Delta x^2}{\hbar} \quad .
\end{equation}
The time scale for broadening needs to be compared with the time needed for
wave propagation over a characteristic distance $L$, e.g., the system size,
\begin{equation}\label{PropagationTime}
t_{\text{prop}} = \frac{mL}{\hbar k} \quad .
\end{equation}
Effects due to dispersion will be important when $t_{\text{broad}}\lesssim 
t_{\text{prop}}$. This condition can be expressed as
\begin{equation}
1\gtrsim \frac{\Delta x^2}{\lambda L} \equiv F,
\end{equation}
where the parameter $F$ is of identical form as the Fresnel parameter that 
distinguishes near-field and far-field diffraction in optics~\cite{BornWolf1999}. A
situation with $F < 1$ is shown in Fig.~\ref{fig:ComboPlot}, where refraction in
spacetime for a Gaussian wave packet scattering off a step potential is shown.
In contrast to Fig.~\ref{fig:Refraction}, the wave packet is seen to broaden
significantly. In addition, the interference fringes are curved and also exhibit
some diffraction in time~\cite{fig2VSfig3Note}. Figure~\ref{fig:ComboPlot} thus
exhibits all three wave phenomena in the spatio-temporal domain: refraction at
the boundary, interference between the incident and reflected components, and
diffraction as the fringes curve. The last of these effects is due to the presence of
a broad range of momentum components in the initial wave packet, causing
considerable broadening  during the scattering event, which is displayed in
spacetime as bending of the interference fringes.

\section{Birefringence in spacetime}
\label{sec:biRef}

Birefringence is an interesting property of some materials, for example calcite
crystals, to exhibit two distinct optical densities, and is routinely used in
undergraduate physics laboratories~\cite{Cloud1973,Camp1996}. Depending on
the polarization of the light, it will move through the crystal in two different paths,
resulting in a double image of the original beam of light. The material itself has two
intrinsic indices of refraction, $n_\parallel$ and $n_\perp$. Here we consider a
spin-1/2 particle in a magnetic field as an analogy. The quantization axis for spin
is most naturally chosen to be parallel to the field direction. Due to the Zeeman 
effect, the particle then experiences a spin-dependent potential $s \mu B$, where 
$s=-(+)1$ for spin-$\uparrow$($\downarrow$) states that are parallel (antiparallel)
to the field direction. This holds regardless of the direction of the external field
relative to the propagation axis for neutral particles~\cite{Neutral}. 
 
\begin{figure}[t]
\begin{center}
\includegraphics[width=8cm]{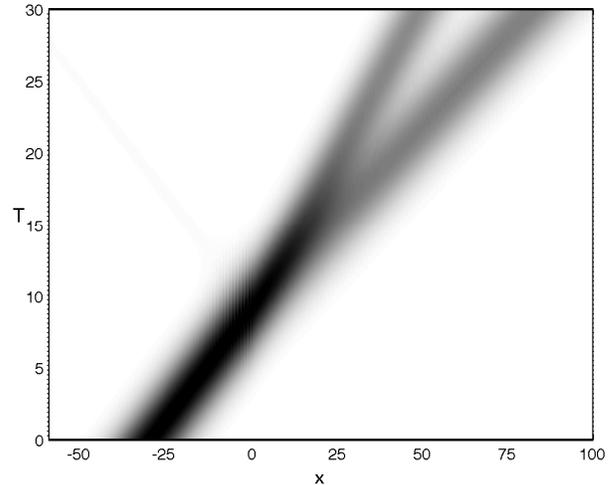}
\caption{\label{fig:Birefringence}
Birefringence in spacetime of a Gaussian wave packet prepared in a field-free region
as a linear superposition of spin-$\uparrow$ and spin-$\downarrow$ components with
equal weight. Parameters are $\Delta k=0.1/\ell$, $x_{0}=-30\,\ell$, $k_{0}=3.5/\ell$,
and $\mu |B|=2.5\,\hbar^2/(m\ell^2)$. The incident wave packet crosses into a region
with finite magnetic field at approximately $T=7.5$. The two distinct paths visible for the
transmitted wave correspond to its separated spin-$\uparrow$ and spin-$\downarrow$
components.}
\end{center}
\end{figure}

In the basis of spin-$\uparrow$/$\downarrow$ eigenstates, the wave function of a
quantum particle is given by a spinor
\begin{equation}
\Psi(x,t)=
\begin{pmatrix}
\Psi_{\uparrow}(x,t) \hfill  \\
\Psi_{\downarrow}(x,t) \hfill 
\end{pmatrix} \nonumber \quad .
\end{equation}
The time-dependent Schr\"{o}dinger equation takes the form
\begin{equation}\label{BirefTDSE}
 i\hbar\frac{\partial}{\partial t} \Psi (x,t) =
-\frac{\hbar^2}{2m}\frac{\partial^2}{\partial x^2} \Psi (x,t) + \hat{H}_{z} \, \Psi (x,t),
\end{equation}
where the effect of the magnetic field is included in the Zeeman matrix
\begin{equation}\label{Zeeman}
\hat{H}_{z}=\begin{pmatrix}
-\mu B & 0 \\
0 & \mu B 
\end{pmatrix} \quad .
\end{equation}
Since the Zeeman term (\ref{Zeeman}) is diagonal, Eq.~(\ref{BirefTDSE}) separates
into two distinct equations for the individual spin components,
\begin{equation}
 i\hbar\frac{\partial}{\partial t}\Psi_{\uparrow/\downarrow} (x,t) =
-\frac{\hbar^2}{2m}\frac{\partial^2}{\partial x^2}\Psi_{\uparrow/\downarrow}(x,t)
\mp\mu B  \Psi_{\uparrow/\downarrow}(x,t) \, ,
\end{equation}
allowing the problem to be treated as the superposition of two independent
propagations. In Fig.~\ref{fig:Birefringence}, the probability density
$|\Psi_{\uparrow}(x,t)|^2+|\Psi_{\downarrow}(x,t)|^2$ is shown as a function of
space and time for a Gaussian wave packet incident from a region with $B=0$
onto a region with $B>0$. The initial spin populations were set equal. As the wave
packet propagates past the discontinuity at $x=0$, it splits up into two parts with
different velocities as the spinor components refract from opposite potential steps
due to the Zeeman splitting. In spacetime, the worldlines of the two spin-components
separate as a result of  the birefringent refraction. Mechanically, this corresponds to
one component being slowed down and the other speeding up after the discontinuity.

It appears from Fig.~\ref{fig:Birefringence} that the wave packets corresponding to
the two opposite-spin components have different widths in region 2 where $B$ is
finite. This phenomenon deserves some brief further discussion. We introduce the
scattering time $\Delta t$ as the time it takes for the wave packet to pass from the
field-free region 1 to region 2. This process of transition can be defined by the
central part of the wave packet moving over the distance $\Delta x$ at the
discontinuity. We then have 
\begin{equation}
\frac{\hbar }{m}k\Delta t_1=\Delta x_1
\end{equation}
for the incoming wave packet. Similarly, in region 2, we have 
\begin{equation}
\frac{\hbar}{m} \sqrt{k^2+\frac{2m\mu B}{\hbar^2}}\,\, \Delta t_{2,\uparrow}=
\Delta x_{2,\uparrow}
\end{equation}
and 
\begin{equation}
\frac{\hbar}{m} \sqrt{k^2-\frac{2m\mu B}{\hbar^2}}\,\,\Delta t_{2,\downarrow}=
\Delta x_{2,\downarrow} \quad .
\end{equation}
Since the scattering time will be the same for all three partial waves, a relation
must hold between incident wave-packet width and transmitted wave-packet
width, which is given by
\begin{equation}
\frac{\Delta x_{2,\uparrow/\downarrow} } {\Delta x_1 } = 
\sqrt{1\pm\frac{\mu B}{\hbar^2k^2}} \quad .
\end{equation}
We thus see that not only is the central velocity affected by the potential step, but
also the width of the wave packet changes during scattering at the interface.
  
 \section{Discussion and conclusions}
 \label{sec:discConc}
 
We point out that the spatio-temporal evolution of a Gaussian wave packet as it
scatters over a potential step represents an analogy to stationary wave refraction.
Inspired by similar phenomena, we have coined the term \emph{refraction in
spacetime\/} for this effect. By this, we have now completed the set of spatio-temporal
analogies for wave phenomena occurring in space. The spatio-temporal equivalent of
Snell's law has been derived and is presented in Eq.~(\ref{eq:genTimeSnell}).

Analogies\cite{BailerJones2009} are commonly used in physics, both as a tool to aid
discovery and in the classroom to compare one situation to another
efficiently~\cite{Treagust1996, Coll2005}. Successful application of an analogy succeeds
in conveying information about an unfamiliar topic, the target,  by relating it to one already
known by the learner, the source. To avoid misconceptions about the target, the limitations
of any utilized analogy must be made explicit. Also, additional understanding can
sometimes be gained by elucidating the limits of an analogy. Our present topic lends itself
to further elaboration on these ideas.

For classical optics and quantum mechanics to be formally analogous in the spatial
domain, the refractive index in the quantum case must be defined by Eq.~(\ref{eq:quantN}).
This sensible choice, however, does not imply that the two theories also behave the same
in spacetime. As it turns out, refraction in spacetime for a quantum wave will behave like its
purely spatial counterpart in the sense that the angle of refraction increases when the index
of refraction decreases. Interestingly, the opposite is true for optical waves in nondispersive
media in the spatio-temporal domain. This different behavior arises as a consequence of the 
differing dispersion properties associated with the two cases because the equivalent of
Snell's law for refraction in spacetime [Eq.~(\ref{eq:genTimeSnell})] involves not only the
refractive indices but also their derivatives with respect to frequency. Conversely, wave
phenomena characterized by refractive indices that have the appropriate frequency
dependence will be formally analogous in both the purely spatial and spatio-temporal
domains. In particular, the refractive behavior in $1+1$ dimensions exhibited by quantum
waves can be faithfully reproduced by electromagnetic waves propagating through a
junction between two waveguides~\cite{note:waveguide}. We suspect that associated
diffraction-in-time effects (as seen in Fig.~\ref{fig:ComboPlot}) will be similarly exhibited in
both systems. Moreover, light trapped in a suitably designed microcavity behaves
identically to matter (quantum) waves~\cite{photonBEC} and can therefore be expected to
exhibit analogous refraction and diffraction phenomena.

Our study of refraction in spacetime further illustrates the general importance of dispersion
for shaping the character of spatio-temporal analogies for wave phenomena. As is
well-known, diffraction in spacetime is absent for electromagnetic waves in vacuum but
becomes possible in a dispersive medium~\cite{Brukner1997}. Alternatively, a spatially
varying refractive index for optical waves results in a Schr{\"o}dinger-type equation in the
beam-propagation direction instead of physical time~\cite{Marte1997}, thus allowing to
explore analogous wave phenomena on yet another level.

\begin{acknowledgments}
We would like to thank an anonymous referee for broadening the scope of our work
to include dispersive optics and steering us toward deriving the spatio-temporal version
of Snell's law in its most general form.
\end{acknowledgments}



\begin{thebibliography}{99}

\bibitem{Brukner1997}
C. Brukner and A. Zeilinger, ``Diffraction of matter waves in
space and time," Phys. Rev. A \textbf{56}, 3804--3824 (1997).

\bibitem{Robinett2006}
R.~W. Robinett, \textit{Quantum Mechanics}, 2nd ed (Oxford University Press, Oxford,
UK, 2006). 

\bibitem{Tannor2007}
D.~J. Tannor, \textit{Introduction to Quantum Mechanics: A Time-Dependent
Perspective} (University Science Books, Sausalito, CA, 2007).

\bibitem{Thaller2000}
B. Thaller, \textit{Visual Quantum Mechanics} (Springer, New York, 2000).

\bibitem{Goldberg1967}
A. Goldberg, H.~M. Schey, and J.~L. Schwartz, ``Computer-generated motion
pictures of one-dimensional quantum-mechanical transmission and reflection
phenomena," Am. J. Phys. \textbf{35}, 177--186 (1967).

\bibitem{Bramhall1970}
M.~H. Bramhall and B.~M. Casper, ``Reflections on a wave packet approach to
quantum mechanical barrier penetration," Am. J. Phys. \textbf{38}, 1136--1145
(1970).

\bibitem{Kiriuscheva1998}
N. Kiriushcheva, S. Kuzmin, ``Scattering of a Gaussian wave packet by a
reflectionless potential," Am. J. Phys. \textbf{66}, 867--872 (1998).

\bibitem{Moshinsky1952}
M. Moshinsky, ``Diffraction in Time," Phys. Rev. \textbf{88}, 625--631 (1952).

\bibitem{delCampo2009}
A. del Campo, G. Garcia-Calderon, J.~G. Muga, ``Quantum transients," Phys. Rpts.
\textbf{476}, 1--50 (2009).

\bibitem{Szriftgiser1996}
P. Szriftgiser, D. GueryOdelin, M. Arndt, J. Dalibard, ``Atomic wave diffraction and
interference using temporal slits," Phys. Rev. Lett. \textbf{77}, 4--7 (1996).

\bibitem{Lindner2005}
F. Lindner, M.~G. Schatzel, H. Wather, A. Baltuska, E. Goulielmakis, F. Krausz,
D.~B. Milosevic, D. Bauer, W. Becker, G.~G. Paulus, ``Attosecond double-slit
experiment," Phys. Rev. Lett. \textbf{95}, 040401-1 -- 040401-4(2005).

\bibitem{birefringNote}
In traditional optics, birefringence occurs when light propagates through a material
that has an index of refraction $n_\parallel$ for the components of light with a given
polarization, and a different one ($n_\perp$) for the perpendicular component. As a
consequence, the wave is split into two components.

\bibitem{Bunge1967}
M. Bunge, ``Analogy in quantum theory: From insight to nonsense," Brit. J. Phil. Sci.
\textbf{18}, 265--286 (1967).

\bibitem{BailerJones2009}
D. Bailer-Jones, \textit{Scientific Models in Philosophy of Science} (University of
Pittsburgh Press, Pittsburgh, PA, 2009).

\bibitem{refIndexNote}
In optics, the refractive index can be defined equivalently in terms of the ratio of
wave numbers or {\em phase velocities\/}: $n_{\text{el}} = k/k_0 \equiv c_0 / c$,
where $c = \omega/k$ is the phase velocity in the medium. Thus vacuum can be
seen as the medium with $n_{\text{el}}=1$, and typical dielectrics have $n_{\text{el}}
>1$.  For quantum mechanics, the dependence on $V_0$ in Eq.~(\ref{eq:quantN})
indicates that the refractive index is measured relative to the level of zero potential
energy. This level, $V_0 = 0$, should most naturally also correspond to the vacuum,
i.e., the case of a free particle that can gain no further kinetic energy. We see from
Eq.~(\ref{eq:quantN}) that $V_0 > 0$ yields $n_{\text{qu}} < 1$, a situation seemingly
unfamiliar from optics. However, cases with $n_{\text{el}} < 1$ are actually
encountered in classical optics for X-rays and are also possible for electromagnetic
waves propagating in plasmas. See, e.g., D.~G. Swanson, \textit{Plasma waves}
(IOP Publishing, Bristol, UK, 2003).

\bibitem{BornWolf1999}
M. Born and E. Wolf, \textit{Principles of Optics: Electromagnetic Theory of Propagation,
Interference and Diffraction of Light} (Cambridge University Press, Cambridge, UK, 1999).

\bibitem{note:phaseVSgroup}
As an interesting aside, we would like to point out that the phase and group velocities
associated with quantum waves show different trends when crossing between regions
of space with different potential. Defining the phase velocity $v_{\text{p,qu}}=\omega/k$
in the usual fashion, we find $v_{\text{p,qu}}=v_{\text{0,qu}} / (2 n_{\text{qu}})$, whereas
the group velocity reads $v_{\text{g,qu}}= v_{\text{0,qu}} n_{\text{qu}}$. Hence, in a case
where the particle (group) velocity increases upon transmission, the phase velocity will
decrease (and \textit{vice versa\/}). This is again a consequence of dispersion. More
details about quantum phase velocity in situations with finite external potential are
discussed, e.g., in L. Bergmann and C. Schaefer, \textit{Optics of Waves and Particles\/}
(de~Gruyter, Berlin, 1999), p.\ 965 and K.~U. Ingard, \textit{Fundamentals of Waves and
Oscillations\/} (Cambridge University Press, Cambridge, UK, 1988), Sec.\ 16.5.

\bibitem{fig2VSfig3Note}
The interference fringes seen in Fig.~\ref{fig:Refraction} are parallel to the
time axis; their only time dependence is a weak modulation due to the
propagating envelopes of the incoming and reflected wavepackets. Hence,
they can be understood as a purely spatial phenomenon. In contrast, fringes in
Fig.~\ref{fig:ComboPlot} exhibit an oscillation in time for fixed $x$. This effect,
which can be attributed to diffraction in spacetime, thus changes the character
of the interference fringes from being spatial in nature to having an additional
temporal dimension.

\bibitem{Cloud1973}
S.~D. Cloud, ``Birefringence experiments for the introductory physics course,"
Am. J. Phys. \textbf{41}, 1184--1188 (1973).

\bibitem{Camp1996}
P.~R. Camp, ``Inexpensive optics for polarized light demonstrations," Am. J. Phys.
\textbf{65}, 449--450 (1997).

\bibitem{Neutral}
For charged particles, there will be additional, velocity dependent, terms due to the
vector potential and the resulting equations will depend on the choice of gauge.

\bibitem{Treagust1996}
D.~F. Treagust, A.~G. Harrison, G.~J. Venville, and Z. Dagher, ``Using an analogical
teaching approach to engender conceptual change," Int. J. Sci. Educ. \textbf{18}, 213--229 (1996).

\bibitem{Coll2005}
R.~K. Coll, B. France, and I. Taylor, ``The role of models/and analogies in
science education: implications from research," Int. J. Sci. Educ. \textbf{27}, 183--198 (2005).

\bibitem{note:waveguide}
The refractive index for classical electromagnetic waves propagating in a waveguide is
given by $n_{\text{wg}} = \sqrt{1 - (\Omega^2/\omega^2)}$, where $\Omega$ is the cut-off
frequency for a particular mode. See, e.g., Sec.~8.3 in J.~D. Jackson, \textit{Classical
Electrodynamics\/}, 3rd ed. (Wiley, Hoboken, NJ, 1999). Specializing Snell's law for
refraction in spacetime [Eq.~(\ref{eq:genTimeSnell})] to this case yields the same
expression in terms of refractive indices as found for quantum waves
[Eq.~(\ref{eq:snellTqu})].

\bibitem{photonBEC}
J. Klaers, J. Schmitt, F. Vewinger, and M. Weitz, ``Bose-Einstein condensation of photons
in an optical microcavity," Nature (London) \textbf{468}, 545--548 (2010). See also the
related perspective by J. Anglin, ``Particles of light," Nature (London) \textbf{468}, 517--518
(2010).

\bibitem{Marte1997}
M. Marte and S. Stenholm,``Paraxial light and atom optics: The optical Schr{\"o}dinger
equation and beyond," Phys. Rev. A \textbf{56}, 2940--2953 (1997).

\end{thebibliography}
\end{document}